# See further upon the giants: Quantifying intellectual lineage in science

Woo Seong Jo[1,2,3], Lu Liu[1,2,3,4], and Dashun Wang[1,2,3,5,*]

[1] Center for Science of Science & Innovation, Northwestern University, Evanston, IL, USA.
[2] Northwestern Institute on Complex Systems, Northwestern University, Evanston, IL, USA.
[3] Kellogg School of Management, Northwestern University, Evanston, IL, USA.
[4] College of Information Sciences and Technology, Pennsylvania State University, University Park, PA, USA.
[5] McCormick School of Engineering, Northwestern University, Evanston, IL, USA
[*] Correspondence to: dashun.wang@northwestern.edu

## Abstract

Newton's centuries-old wisdom of standing on the shoulders of giants raises a crucial yet underexplored question: Out of all the prior works cited by a discovery, which one is its giant? Here, we develop a discipline-independent method to identify the giant for any individual paper, allowing us to better understand the role and characteristics of giants in science. We find that across disciplines, about 95% of papers appear to stand on the shoulders of giants, yet the weight of scientific progress rests on relatively few shoulders. Defining a new measure of giant index, we find that, while papers with high citations are more likely to be giants, for papers with the same citations, their giant index sharply predicts a paper's future impact and prize-winning probabilities. Giants tend to originate from both small and large teams, being either highly disruptive or highly developmental. Papers that did not have a giant tend to do poorly on average, yet interestingly, if such papers later became a giant for other papers, they tend to be home-run papers that are highly disruptive to science. Given the crucial importance of citation-based measures in science, the developed concept of giants may offer a useful dimension in assessing scientific impact that goes beyond sheer citation counts.

# 1 Introduction

"If I have seen further, it is by standing on the shoulders of giants," Isaac Newton's famous 1675 letter to Robert Hooke highlights a fundamental feature of science: its cumulative nature. Indeed, new insights and discoveries rarely emerge in isolation; instead they build on prior scientific work. Since scientists throughout the ages and across disciplines all acknowledge ideas that inspired their research, we have an opportunity to explore citation relationships to better understand how new research makes use of influential work (Clauset, Larremore, & Sinatra, 2017; Fortunato et al., 2018; Garfield, 2006; Price, 1965; Radicchi, Fortunato, & Castellano, 2008; Redner, 2005; Waltman, 2016; Wang & Barabási, 2021; Wang, Song, & Barabási, 2013). Here, we hone in on a specific, underexplored question: Is there a way to estimate, given any paper, which reference is the "giant" whose shoulders the new research stands upon?

Here we take advantage of the citation relationships between papers and develop a network-based method that aims to estimate the relative intellectual significance of each reference to a paper, allowing us to estimate the potential giant for a paper. Our leading hypothesis is that, while each paper has many references, their importance to the paper can be uneven, and the relationships among the listed references, revealed through the overall citation network, may help us understand the intellectual significance of each reference in the context of the overall scientific discourse.

Citations are essential to scientific communication, allowing scientists to condense knowledge, bolster the strength of their evidence, attribute prior ideas to appropriate sources, and more (Wang & Barabási, 2020). Partly because canonical papers tend to inspire follow-up research which builds on them, citation relationships have been widely used to quantify scientific impact (Bergstrom, West, & Wiseman, 2008; Cole & Cole, 1974; Garfield, 2006; Hirsch, 2005; King, 2004; L. Liu et al., 2018; Radicchi et al., 2008; Sinatra, Wang, Deville, Song, & Barabási, 2016; Uzzi, Mukherjee, Stringer, & Jones, 2013; Waltman, 2016; Wang et al., 2013; Way, Morgan, Larremore, & Clauset, 2019; Wu, Wang, & Evans, 2019; Wuchty, Jones, & Uzzi, 2007). At the same time, citations can be affected by myriad factors: publication venue, year, field of study, among many other reasons why authors cite a given paper (Aksnes, 2006; Moravcsik & Murugesan, 1975; Radicchi, 2012; Simkin & Roychowdhury, 2002), contributing to noise in evaluating and comparing their relative importance.

The need to quantify scientific impact by considering citations of different importance has inspired various methods to identify the key references. Some examined the citation context by analyzing the number of mentions and the sections and relative positions that a reference appears (Bornmann & Daniel, 2008; Boyack, van Eck, Colavizza, & Waltman, 2018; Ding, Liu, Guo, & Cronin, 2013; Ding et al., 2014; Hu, Chen, & Liu, 2013; Jones & Hanney, 2016). Some combined features from the full text, citation counts and abstract similarity, and trained a classifier to predict important references (Hassan, Akram, & Haddawy, 2017; Zhu, Turney, Lemire, & Vellino, 2015). Some developed local diffusion method on the citation network and ranked references by their diffusion score (Cui, Zeng, Fan, & Di, 2020). Other studies conducted surveys for authors and asked them to identify references that shaped the research idea and influenced the research (Tahamtan & Bornmann, 2018; Zhu et al., 2015). These studies have also contributed to the development of new metrics, such as weighting citations by the number of mentions and the diffusion score on the citation network.

Here we study 33 million papers indexed by the Web of Science (WOS) between 1955 and 2014, and 962 million citations among them (SM S1). To quantify the intellectual significance of each reference to a given paper, we first use co-citation relationships to establish a measure of proximity between references, which measures how many times two papers are cited together by other papers. Co-citation has been used for a variety of purposes from quantifying the topical relevance among papers (Chen, 2006; S. Liu & Chen, 2012; Small, 1973) to author credit allocation within a paper (Shen & Barabási, 2014) to evaluating impacts of authors (Ding, Yan, Frazho, & Caverlee, 2009) and suggesting relevant references (Sarol, Liu, & Schneider, 2018). To identify the "giant" for a given paper, we first take all the papers ever published up to the publication year of the focal paper, and identify all co-citation relationships (Figure 1A), with link weight indicating the number of times two papers are co-cited. This co-citation network can act as a proxy for the overall knowledge space at the time (Figure 1B). Next, for the focal paper, we locate all its references to obtain the reference subnetwork embedded in the overall co-citation network, approximating the intellectual context in which the paper is placed (Figure 1C). Here we focus on papers that contain at least five references to ensure we have enough nodes for each reference subnetwork, resulting in 25M papers in total.

To identify local significance of a reference within the subnetwork while taking into account its overall influence in the global network (Lü et al., 2016; Lü & Zhou, 2011), we develop a new method by borrowing concepts from democratic voting and percolation theory. Our key insight is to give each reference the same number of "votes" to link to their most relevant paper in the global co-citation network, but keep only the links formed within the reference subnetwork (only counting the votes within the specific context). We hypothesize that the giant of a paper should appear in its reference list and have high topical relevance to the body of work, prompting us to focus on the votes in the reference subnetwork. At the same time, the giant paper may also be captured by collective recognition within a scientific community. To this end, our method recognizes the collective nature of the intellectual lineage by exploring rich information embedded in the global co-citation network. Specifically, we begin by adding one connection (a vote) for each reference, and monitor the connectivity within the subnetwork (Figure 1D). We increase the number of connections iteratively, and stop when the reference subnetwork starts to coalesce into a cluster, suggesting plausible knowledge structures beginning to emerge within this subnetwork. Resorting to percolation theory, we stop at the minimal n votes needed for the subnetwork to cross the percolation threshold for the corresponding random network (Newman, 2010), i.e., $\langle k_n > 1 \rangle$, where references within the subnetwork have at least one connection on average (Figure 1E). We then select the node with the largest degree ($k$) as the giant paper (SM S2). The idea is to approximate the local importance of a reference by counting the 'votes' it received from other references. And the one with the most 'votes' suggests its importance among other references. At the same time, the idea of percolation threshold also affords us the possibility of not always identifying a giant for a paper. Indeed, if, however, the reference subnetwork remains completely isolated at $n = 1$ (Figure 1F), suggesting that the paper's references all belong to distinctive parts of the overall knowledge space, we stop our algorithm at $n = 1$ without identifying giants for such cases.

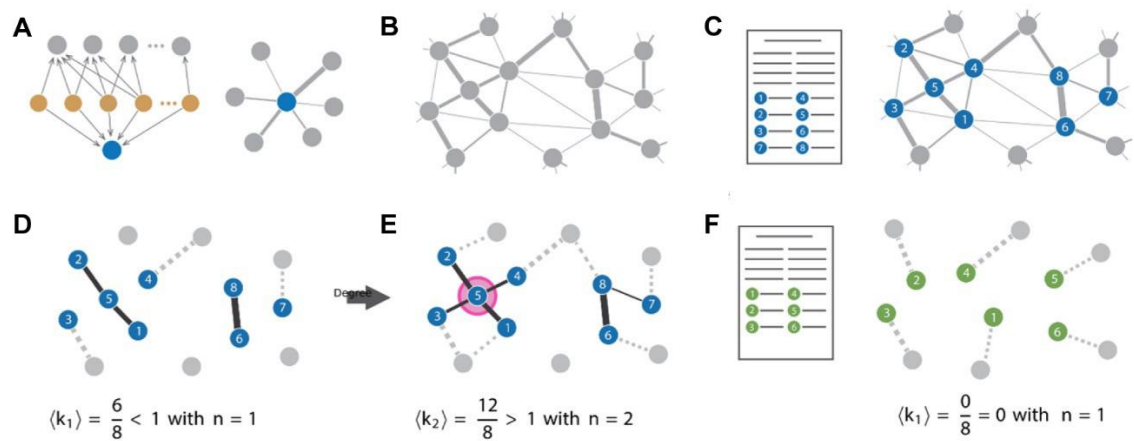

**Figure 1.** Identifying the giant paper. (A) For a given reference (blue), we find all other papers that are co-cited with this reference (co-cited papers in grey) with link weight indicating the number of co-citations. (B) We can follow the procedure in (A) to construct the entire co-citation network for all papers published up to a certain year, representing the overall knowledge space at the time. (C) To identify the giant for a given paper, we use the co-citation network in its publication year, and locate all its references (blue nodes) to pinpoint the reference subnetwork from the overall co-citation network. (D) We give each reference the same number of "votes" to link to their most relevant paper in the global co-citation network (both solid and dashed lines), but keep only the links formed within the reference subnetwork (solid line). We begin by adding one connection (a vote) for each reference, and monitor the connectivity within the subnetwork. (E) We increase the number of connections iteratively, and stop when the average degree of the reference subnetwork is greater than 1, minimum requirement to have a cluster. We then pick the node with the largest degree as the giant paper (highlighted in purple). (F) If, however, the reference subnetwork remains completely isolated at $n = 1$, we stop our algorithm without identifying giants for such cases.

One advantage of our method is to overcome mixed signals introduced by the skewed citation distribution (Barabási & Albert, 1999; Cui et al., 2020; Price, 1965; Radicchi et al., 2008). Indeed, since highly cited papers tend to dominate the subnetwork of references, directly applying existing network-based methods to the co-citation network tends to favor papers with an overall high citation counts, which may or may not be the specific giant to a paper. One can somewhat mitigate this issue by keeping for example the most essential links using network sparsification methods such as backbone extraction (Radicchi, Ramasco, & Fortunato, 2011; Serrano, Boguñá, & Vespignani, 2009), yet it appears insufficient to overcome the dominance of highly cited papers. This highlights one of the many challenges of this task: to identify the giant for a specific paper, we need to balance between a reference's local importance to the specific context and its overall scientific impact. At the same time, prior work reveals that creative works may not necessarily build from existing literature (Tahamtan & Bornmann, 2018), and our method allows us to identify papers without giants, which is difficult to achieve using citation count or the backbone algorithm. Overall, we recognize that high-citation papers are important and often instrumental in the development of the field, but it does not guarantee that they are the most relevant for a specific paper in a field. This prompts us to take into account both the local relevance of a paper within the reference list and its global impact in the overall citation network, which may offer a complementary dimension of a paper's impact to its citation count. We also note that each of the steps in the method can be further refined with different stopping criteria and different number of giants to pick for a paper. Here we choose this specification of the method for its simplicity and leave for future work to systematically study the different variants of the method, which we shall discuss in more detail in the discussion and limitation section of the paper.

## 2 Results

We apply our method to millions of papers published over the past 60 years, allowing us to study the giants we identify and examine patterns of intellectual lineage in science. First, we find that about 95% of papers have a giant associated with them (Figure 2A). This overwhelming proportion of papers with identifiable giants offers quantitative evidence for the cumulative nature of science. We further find that, despite the exponential growth of science and rise of interdisciplinary research and collaborations, this prevalence of giants has held true over the ages but has risen gradually over time (Figure 2B), growing from 91.6% in 1955 to 95.8% in 2014, which implies that research today is increasingly conducted on the shoulders of giants. Yet at the same time, the papers that serve as giants are remarkably concentrated, since only 12% of the papers we analyzed fell into this category (Figure 2A), consistent with what the literature suggests (Bornmann, de Moya Anegón, & Leydesdorff, 2010; Cole & Cole, 1972). We further find that these results are remarkably consistent across the disciplines we studied (Figure 2C-D). Hence despite the inherent differences in norm and culture across disciplines, the shoulders of giants appear universally appreciated. Interestingly, of the 12 different scientific fields, multidisciplinary science has the highest fraction of giant papers (15.4%, Figure 2D), suggesting a premium in bringing together diverse approaches. Overall, these results suggest that while the vast majority of papers stand on the shoulders of giants, the weight of scientific progress rests on only a few shoulders.

Is the identified giant for a given paper the most cited within its references? For each paper that has a giant, we calculate the citation counts of all its references when the paper was published. We find that the vast majority of giant papers are *not* the most cited within the reference list (72.5%, dashed line in Figure 2E), and this fraction has mostly been trending upward over the past 60 years (Figure 2E). For instance, in 1955, 44% of giants were also the most cited references, but that number decreased to only 26% in 2014. These results suggest being the most-cited does not guarantee to be the giant, implying that our measures of giant may capture complementary dimensions of impact to citation counts, which prompt us to further examine the overall characteristics of a giant paper.

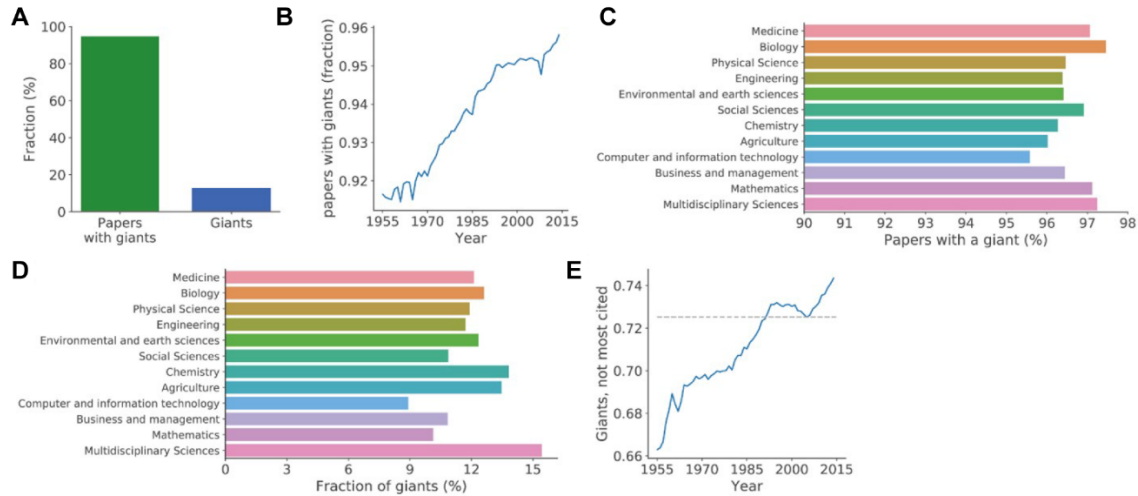

**Figure 2.** Prevalence of giants in science. (A) Out of the 25M papers we studied, 95% of them have a giant (green). By contrast, only 12% of papers later become giants for other papers (blue). (B) The fraction of papers that have an associated giant increases over years. (C) Breaking down the fraction of papers with giants by 12 different fields. (D) Breaking down the fraction of papers that later become a giant by 12 different fields. Papers in multidisciplinary sciences have the highest likelihood to become a giant. (E) We calculate the fraction of papers whose giant is not the most cited paper within its reference list, and show this fraction over time.

To understand what kinds of papers tend to become a giant, we introduce a new index: the giant index ($G$), which calculates the number of times a paper is a giant for other papers. We then compare a paper's $G$ with its citation count ($C$). We first find that the likelihood of being a giant posts a highly non-linear relationship with its citation counts. Indeed, among papers with exceptional impact ($C > 1000$), virtually all of them have a positive giant index. The probability for a paper to be a giant $P(G > 0)$ increases with citation counts but undergoes a sharp crossover for papers with respectable but more moderate citations (Figure 3A). To ensure the observed trend is not affected by self-citations, we remove self-citations and repeat our analysis, arriving at the same conclusion (SM S3). These results suggest that not all papers can be a giant, and a paper's potential to become a giant may be related to its citation impact, prompting us to further examine the correlation between a paper's giant index and its citations (Figure 3B). We find that for papers with dozens to hundreds of citations, their giant index and citations follows a super-linear relationship, suggesting that an increase in a paper's citations is associated with an increasing return in the rate at which a paper becomes a giant. Yet for papers in the right tail of citation impacts, their giant index roughly follows a linear relationship with their citations, consistent with the trend shown in Figure 3A, suggesting that papers with exceptional impact are disproportionately more likely to be the shoulders that carry scientific progress in their field.

At the same time, the overall correlation between the giant index and citations (Figure 3B) also masks heterogeneous relationships between the two. Indeed, since both $P(G)$ and $P(C)$ follows a fat-tailed distribution (SM S4) and by design $G \leq C$, we calculate the conditional probability $P(G|C)$ (Figure 3C). We find that, as $C$ increases, the distribution systematically shifts to the right, consistent with the correlations observed in Figure 3A-B. Yet, for papers with the same level of citations, their giant indexes are still characterized by a high degree of heterogeneity, suggesting that a high citation count does not necessarily guarantee a high giant index. Next, we show, this discrepancy between the giant index and citation counts offers signals for a paper's future impact.

We select papers published in *Physical Review Letters* (*PRL*) between 1990 and 2000 but have a similar level of citation impact after five years of publication (within the range of $100 \leq C_5 \leq 200$ by Year 5) (Figure 3D). We divide these papers into three groups based on their giant index at year 5 ($G_5$): high $G$ group (Top 10% in giant score $G_5$, $\langle G_5 \rangle \approx 31.3$), low $G$ group (bottom 10% in $G_5$, $\langle G_5 \rangle \approx 1.0$), and non-giant group ($G_5 = 0$). We then trace the citations of these papers over the next ten years. We find that at year 5, the three groups follow a similar citation distribution, by construction. Yet with time, the high $G$ group clearly stood out from the pack, collecting citations at a much faster rate than the other two groups. Interestingly, there is a statistically significant difference between the $G \approx 1$ and $G = 0$ group by year 15, suggesting that papers with a small giant index is likely to have higher future impact. We repeat the same analysis for a multidisciplinary journal, selecting papers published in *Proceedings of the National Academy of Sciences* (*PNAS*) ($100 \leq C_5 \leq 200$, Figure 3E), finding again the same patterns. Taken together, Figure 3 shows that while a paper's giant index and citation counts are overall correlated, papers with the same citations can have vastly different giant indexes, and that difference in giant index appears to substantially reveal a paper's potential for future impact. These results suggest that giant index offers additional information on a paper's role in science that goes beyond its citation counts. Together, they reflect the idea that not all citations are the same, and those that frequently lend their shoulders to others tend to distinguish themselves from those that do not.

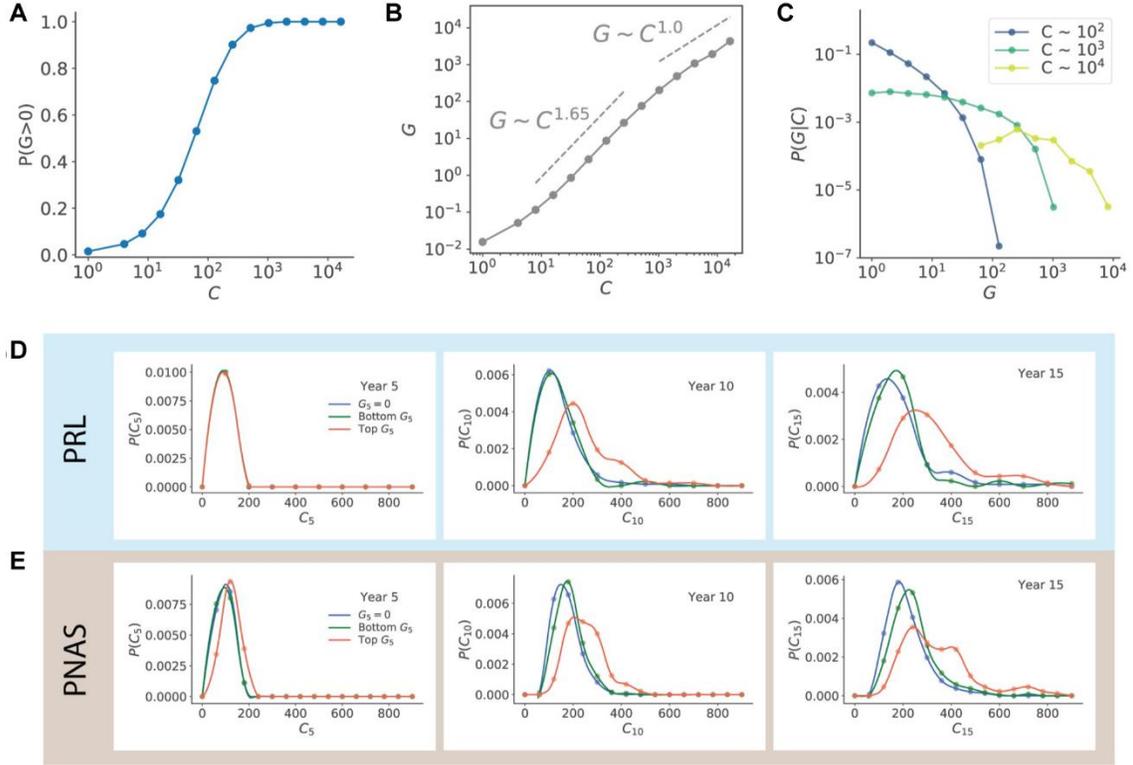

**Figure 3.** The giant index. (A) The fraction of papers with a non-zero giant index as a function of citations. (B) The giant index $G$ as a function of $C$. The two quantities follow a super-linear relationship when $C < 1000$ but exhibit a linear relationship for high $C$ regime ($C > 1000$). (C) Conditional distribution $P(G|C)$ shows that papers are characterized by a heterogeneous giant index even though they have the same level of citation impacts. (D) We pick papers published in *PRL* between 1990 and 2000 that have between 100 and 200 citations in 5 years ($100 \leq C_5 \leq 200$). Depending on their giant index at year 5 $G_5$, these papers exhibit different levels of future impacts. (E) The same analysis as (D) but for a group of papers published in *PNAS* ($100 \leq C_5 \leq 200$).

To understand potential forces that might facilitate the production of giants, we further examine the organization of scientific activity, probing the role of teams in shouldering scientific progress. Indeed, research shows that small and large teams are differentially positioned for innovation (Wu et al., 2019): large teams tend to excel at furthering existing ideas and design, whereas small teams tend to disrupt current ways of thinking with new ideas and opportunities. This distinction prompts us to measure if the likelihood of producing giants vary by team size. To control the effect of field and time, we normalize citations and giant indexes with the average value for papers published in the same field and year (Radicchi et al., 2008), computing the normalized citation $C/\langle C \rangle_{f,y}$, and normalized giant index $G/\langle G \rangle_{f,y}$. We first find that, across different fields, the normalized citation, increases with team size $M$ (Figure 4A), confirming previous studies showing the citation premium conferred upon large teams (Wu et al., 2019; Wuchty et al., 2007). We then repeat this analysis for giant papers ($G > 0$), finding that the normalized giant index posts a U-shaped curve (Figure 4B), and this nonlinear relationship holds the same across different fields (shaded curves in Figure 4b). These results suggest that while works by large teams tend to garner higher citations, giants who frequently lend their shoulders tend to originate from both small and large teams.

The relationship between team size and research outcomes prompted us to ask if the observed nonlinear relationship between giant index and team size is related to the character of work that teams of different size produce. Here, we measure the relationship between a paper's giant index and its disruption percentile $DP$. We calculate the disruption score following prior work (Funk & Owen-Smith, 2017; Wu et al., 2019). For each paper, we calculate the number of subsequent papers citing the paper but not its references ($n_i$), the number of subsequent papers citing both the paper and its references ($n_j$), and the number of subsequent papers citing the references but not the paper ($n_k$). The raw disruption score ($D$) is defined as $D = \frac{n_i - n_j}{n_i + n_j + n_k}$. We further normalized the raw score to its percentile, the disruption percentile ($DP$). $DP$ measures the relative ranking among all papers published in the same year, with 100 indicating the most disruptive and 0 as the most developmental. We find that giant index sharply increases for both highly developmental ($DP < 20$) and highly disruptive ($DP > 80$) work (Figure 4c). These results illustrate the divergent characters of giants in science, being either highly disruptive or highly developmental. Both types of giants shoulder scientific progress, but they move science forward in different ways.

One advantage of our method is to allow the flexibility of not identifying a giant for a paper. Indeed, as discussed in Figure 1, if all the references of a paper are initially isolated in the reference subnetwork, it suggests that the paper draws upon disparate rather than established knowledge clusters, and for these papers, our method proceeds without assigning a giant (i.e., papers without giants). Of the 25M papers we studied, 5.2% of them fall into this category. Figure 4D plots the citation distributions for papers with and without a giant, respectively. We find that papers without a giant tend to have overall fewer citations than those that stand upon a giant's shoulder, further suggesting the importance of giants in the production of knowledge. Yet at the same time, Figure 4D also reveals an intriguing observation: even among papers without a giant, their citations are characterized by a high degree of heterogeneity, indicating that some papers do eventually garner high impacts, albeit uncommonly. Indeed, we find that, among the papers without giants, 10% of them go on to become giants for others. We separate the papers without giants into two groups, calculating the relative probability of observing a $G > 0$ vs $G = 0$ paper as a function of citations, and find that the two groups follow clearly divergent patterns (Figure 4E): If a paper neither became a giant for others ($G = 0$) nor stood on one's shoulder when published, its impact is mostly concentrated within the low citation region, and the probability for such papers to garner higher citations diminishes rather rapidly. By contrast, those that did not have a specific shoulder to rely upon but later became a giant for others are systematically overrepresented in the high citation region, as their relative abundance rises precipitously with citations. These results paint a highly polarized view for papers without a giant. On the one hand, such papers have a rather limited impact on average, suggesting that skipping the shoulders substantially limits a paper's ability to "see further". Yet on the other hand, perhaps counterintuitively, papers without a giant may also become home-run papers with right-tail citation impact. One conjecture is that papers that emerge from seeming vacuum that lies between knowledge clusters may reorient the existing knowledge in a way that offers new ideas and opportunities. To test this conjecture, we measure the disruption percentile $DP$, and find that papers without giants are sharply overrepresented in the highly disruptive region (Figure 4F), and much more likely to be produced by small teams (Figure 4F inset).

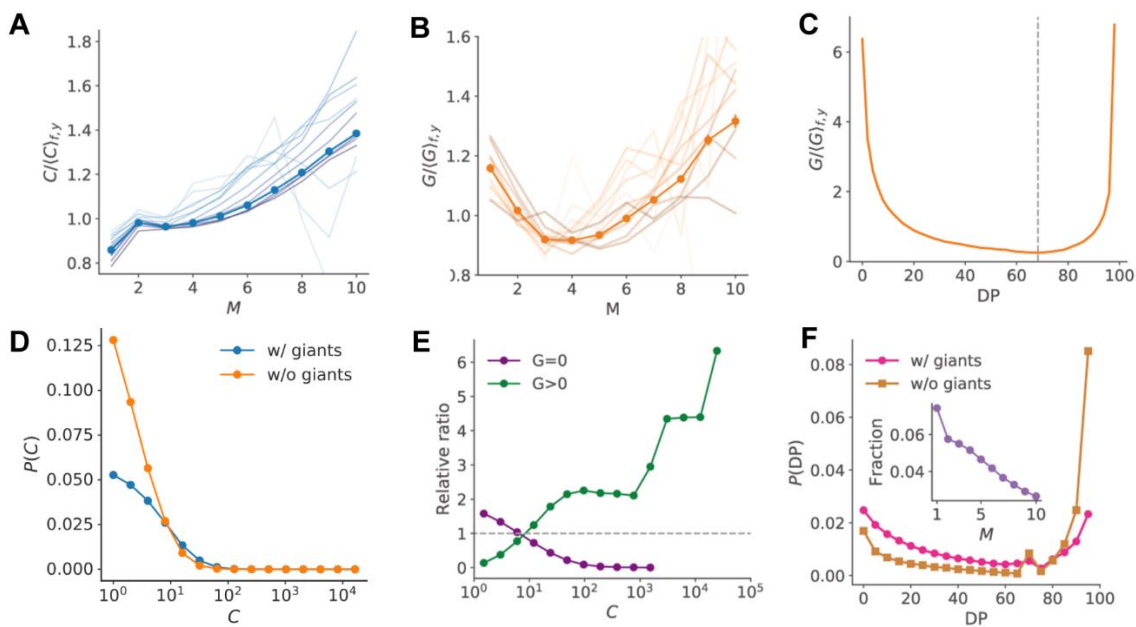

**Figure 4**. Teams, disruption, and giant index. (A) We normalize citation by field and year $C/\langle C \rangle_{f,y}$, and plot it as a function of team size $M$. (B) We normalize the giant index by field and year, finding that it follows a U-shaped curve with team size. (C) The normalized giant index as a function of disruption percentile $DP$, showing that either highly developmental ($DP \leq 20$) or highly disruptive ($DP \geq 80$) papers tend to a high giant index. (D) Distributions of citations for papers with or without a giant. (E) We **categorize** papers without giants into two groups based on their giant index: $G = 0$ vs $G > 0$. The two groups show clearly different relationships with the citation impacts of these papers. (F) The distribution of disruption percentile $DP$ for papers with (red circle) or without (yellow squares) a giant, showing that giants without a giant are disproportionately more likely to be a highly disruptive paper. (Inset) The fraction of papers without a giant as a function of team size, showing such papers are more likely to be produced by small teams.

Lastly, as a further validation, we show that the giant index offers a simple yet additional early signal for the Noble prize-winning papers. As the most prestigious prize in science, the Nobel Prize recognizes some of the most crucial scientific breakthroughs. There have been constant attempts in identifying Nobel prize-winning discoveries based on citation counts (Garfield & Malin, 1968; Revesz, 2015; Zakhlebin & Horvát, 2017). Despite its occasional success, citations appear to be a noisy signal for the Nobel, due to a simple reason: while Nobel prize-winning papers all tend to be highly cited, having high citations does not guarantee a Nobel. This raises an intriguing question: could the giant index offer additional information in differentiating the prize-winning papers beyond citation counts? To answer this question, we identified 370 Nobel prize-winning papers for the physics, chemistry and medicine Nobel awarded between 1955 and 2014, and compared their giant index with papers published in the same year and field with similar citations (the comparison group). We find that, by construction, the citation distribution $P(C)$ is largely indistinguishable between the two groups of papers (Figure 5A, inset), yet the prevalence of Nobel prize-winning papers systematically increases with their giant index (Figure 5A). Indeed, if we just compare the giant index of prize-winning papers with that of their non-prize-winning counterparts, we find a majority of prize-winning papers have a higher $G$ (67%) (Figure 5B). Hence even though the two groups have the same citations, simply comparing their giant index offers stronger signal to distinguish them. Indeed, we further compared the median giant index between the two samples, finding that across physics, chemistry, and medicine, the median giant index for prize-winning papers is more than twice of that of the control group (Figure 5C). Note that the exercise shown in Figure 5 illustrates a relatively simple approach, suggesting the utility of the giant index can be further improved with additional features and more sophisticated models. Together, Figure 5 not only offers further evidence that incorporating the giant index may offer additional signal to identify influential work than citation count alone; it also suggests distinctions between getting cited and being the reliable shoulder for ensuing science.

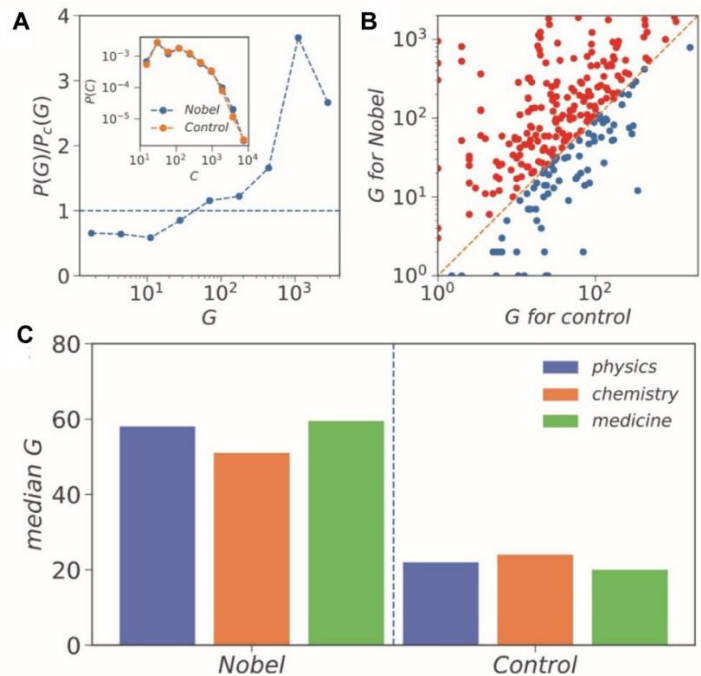

**Figure 5.** Giant index and Nobel prize-winning papers. (A) For each Nobel prize-winning paper, we construct a comparison set of papers that are published in the same year and field and have a similar number of citations (inset). We then compare the distributions of giant index for these two groups of papers, calculating the relative probability $P(G)/P_C(G)$ as a function of the giant index, where $P(G)$ is the distribution of giant index for Nobel prize papers and $P_C(G)$ is the distribution of giant index for their non-prize-winning counterpart. (B) We compare the giant index between the prize-winning papers and their non-prize-winning counterparts in the comparison set. Red dots indicate prize-winning papers have a higher G than those in the comparison group. (C) The median giant index for the Nobel prize-winning papers is 58 in physics, 51 in chemistry, and 59.5 in medicine, respectively. By contrast, the median giant index for papers in the comparison group with similar level of citations is 20 in physics, 22 in chemistry and 24 in medicine, respectively.

## 3 Discussion

In summary, here we present a quantitative framework to test Newton's canonical insights on standing on the shoulders of giants. Several past efforts have attempted to identify a paper's influential references. While these attempts are mostly limited to small samples and require domain knowledge or manual processes for classification, our method offers an alternative way to quantitatively identify crucial references within a paper, which has several advantages that are worth noting. First, the method is discipline or journal independent, easily applicable to any paper corpus where citation information is available. Second, it does not require ex post information, meaning that the giant paper can be identified at the time of publication, which increases the practical utility of the method. Third, as a framework, the method can be further extended to allow for more flexibilities. For example, although the proposed method places an implicit assumption that each paper has just one giant, it can be easily extended to incorporate multiple giants by picking the top $k$ references of a paper or references whose weight in the co-citing network is above a threshold. Indeed, it is reasonable to assume that a paper may build upon multiple giants. Understanding papers with multiple giants is an important direction for future work. It would be interesting to analyze these cases and compare with results obtained by picking one giant, which could be especially insightful when it comes to multidisciplinary work and could further illuminate the roles of giants in the development of new ideas.

The overall intuition behind the proposed measure is to recognize the uneven intellectual influence of the references to a specific paper by quantifying the relationships among the references cited in the paper through co-citation networks. As such, the method offers one way to balance between a reference's local importance to the specific context and its overall scientific impact, helping us better understand the intellectual significance of each reference in the context of the overall scientific discourse. Nevertheless, as an initial attempt to understand the role and characteristics of giants in science, there are several important limitations of our work, each suggesting important directions for future work. First and foremost is the validation of the proposed framework. In this paper, we calculated the giant score for Nobel prize-winning papers. Future work may compare the giant identified by our method to existing metrics, such as the number of mentions and the appearance in difference sections of papers through full text analysis. One could also assess the validity of the giant index through surveys, by asking for example the lead authors of a paper to identify the most important reference for their work, as well as asking whether there was any giant at all. This direction is especially important given the algorithmic nature of our method, which may represent a crude approximation for a paper's intellectual lineage which inherently depends on a range of social and institutional factors. Second, while our paper proposes one way to quantify the intellectual lineage among papers, there could be potentially several other ways to quantify the shoulders of giants in science, representing fruitful directions for future work, which may lead to more robust methods and further insights. It is also important for future work to compare the performance and validity of different methods in identifying giants. Third, it is important to keep in mind that, as with many citation-based indicators, the giant index does not account for the many individual and institutional factors influencing a paper's future impact. Further, citation-based measures may also have inherent biases against recent work, as it takes time for citations to accumulate, suggesting that it may be more difficult to identify the giant(s) if a paper builds on a more recent body of work. Lastly, the prediction task in the paper offers correlational evidence supporting the relevance of the giant index. Future work with causal design may help improve the causative interpretation of the idea of standing on the shoulders of giants.

Overall, given the crucial importance of citations in science decision making, including hiring, promotion, granting and rewards, the developed concept of giant and its associated giant index may offer a useful dimension in our quantitative understanding of science by allowing us to appreciate those who shoulder the scientific progress. As such, this measure is not limited to individual discoveries, but offers a complementary dimension to the growing literature of the science of science (Fortunato et al., 2018; Wang & Barabási, 2020), and can be fruitfully applied to assess the role of giants in careers, teams, institutions, and more, pointing to promising future directions.

**Author Contributions.** Woo Seong Jo: Data curation, Formal Analysis, Investigation, Methodology, Visualization, Writing–original draft, Writing–review & editing. Lu Liu: Formal Analysis, Investigation, Methodology, Visualization, Writing–original draft, Writing–review & editing. Dashun Wang: Conceptualization, Funding acquisition, Investigation, Methodology, Project administration, Supervision, Writing–original draft, Writing–review & editing.

**Competing Interests.** The authors declare no competing interests.

**Data availability**. The Web of Science data are available via Clarivate Analytics. The Nobel Prize data are from https://www.nature.com/articles/s41597-019-0033-6.

**Funding Information.** This work is supported by Air Force Office of Scientific Research under award nos. FA9550-15-1-0162, FA9550-17-1-0089, and FA9550-19-1-0354.

**Acknowledgments**. The authors thank all members of the Center for Science of Science and Innovation (CSSI) for invaluable comments.

# Supplementary Materials for See further upon the giants: Quantifying intellectual lineage in science

Woo Seong Jo[1,2,3], Lu Liu[1,2,3,4], and Dashun Wang[1,2,3,5,*]

[1] Center for Science of Science & Innovation, Northwestern University, Evanston, IL, USA.
[2] Northwestern Institute on Complex Systems, Northwestern University, Evanston, IL, USA.
[3] Kellogg School of Management, Northwestern University, Evanston, IL, USA.
[4] College of Information Sciences and Technology, Pennsylvania State University, University Park, PA, USA.
[5] McCormick School of Engineering, Northwestern University, Evanston, IL, USA
[*]Correspondence to: dashun.wang@northwestern.edu

## S1. Data Description

In this project, we collected 33 million publications across 216 disciplines together with 962 million citation links among them, constituting all publication records indexed in the Web of Science (WOS) database between 1955 and 2014. WOS provides for each publication detailed information such as the unique id, title, journal, year, subject and publication type (e.g. article, review, editorial materials, news). We focus on research outputs and analyze records with publication type as articles and letters. We further filter out papers with fewer than 5 references to ensure that we have enough nodes for the reference subnetwork of each paper. In sum, we have 25 million papers in total for the analysis.

## S2. Method for identifying the giant

**Construct reference subnetwork for each paper.** We first construct for each year a global co-citation network among all papers published up to the year. Two papers are connected if they are co-cited at least once, with the link weight indicating the number of times two papers are co-cited. This co-citation network captures the overall knowledge space at the time. Then for a focal paper p published in this year, we focus on its references and borrow the concept of democratic voting to give each reference the same number of 'votes' to identify its most relevant paper in the global co-citation network. This allows us to construct for each focal paper a subnetwork among references, with link between a reference $r_i$ to another reference $r_j$ indicating $r_j$ has the highest weight (mostly co-cited) among all neighbors of $r_i$ in the global co-citation network. If the most co-cited paper of $r_i$ is not among the paper p's references, we don't connect $r_i$ to any other references. We increase the number of connections iteratively from the most co-cited paper to the second, third until the top n, when the average degree $\langle k_n \rangle$ of the reference subnetwork is above the percolation threshold for the corresponding random network ($\langle k_n \rangle > 1$), which captures the minimal *n* votes needed for the subnetwork to form a cluster of nodes. However, if there is no connection between any of the references when $n = 1$, we assume that the paper's references all belong to distinctive parts of the overall knowledge space and stop identifying its giant for such cases.

**Identifying the giant in each reference subnetwork.** We quantify the importance of each reference to the focal paper by the total number of links formed within the subnetwork, and assign the giant as the reference with the maximum number of links $k_{max}$ in the subnetwork. If multiple papers have $k_{max}$ links, we further compare their co-citation counts and pick the one with the highest weight. Note that we can calculate an importance score $s_i$ for any node *i* in subnetwork following the function

$$s_i = \frac{k_i}{k_{max}} \frac{w_i}{w_{k_i,max}} f(i)$$

where $k_i$ is the number of links, $w_i$ is the weight, $w_{k_i,max}$ is the maximum weights for nodes having $k\_i$ links, and $f(i)$ is an damping function. In this paper we set $f(i) = \delta(k_i - k_{max}, w_i - w_{max})$ to assign 1 for the giant and 0 for the rest. We can modify this equation to account for the relative importance of any node in the subnetwork and identify multiple giants if needed, showing the flexibility of the overall framework.

**Case study on CRISPR-Cas9.** In this section we conduct case and visualize the reference subnetwork for the paper on CRISPR-Cas9 (Jinek et al., 2012) (Figure S1). Our algorithm identifies Barrangou et al. (Barrangou et al., 2007) as the giant (the darkest blue node), which is considered a breakthrough piece on CRISPR study. Furthermore, the most cited reference on microRNA (Lewis, Burge, & Bartel, 2005) in 2005 is not selected as the giant under our method. Rather, it is isolated from any other nodes, suggesting its impact is largely indirect to the CRISPR-Cas9 discovery compared to other references despite its high citation count. This case study suggests that our method offers complementary information to the citation measures in different disciplines.

## S3. Robustness check: self-citations

To ensure that our definition of giants is not affected by self-citations, we exclude self-citations in our method and ignore the giant of a focal paper if they were written by the same author(s). Specifically, if the focal paper shares at least one author with identical first initial and the last name with its giant, we consider it a self-citation and exclude it in calculating the giant score $G$ for the giant. This method offers a conservative estimation on the effect of self-citation, as it may over-estimate self-citations especially for giant papers written by authors with common names. We then examine the fraction of giants measured with and without self-citation for papers with different citation counts (Figure S2). We find that although the fraction of giant becomes slightly smaller for low-citation papers, in general the self-citation has limited impact on a paper's likelihood of becoming a giant, especially for influential papers.

## S4. Properties of giant papers

**Citation of giant and non-giant papers.** We compare the distribution of citations $P(C)$ for giant papers ($G > 0$) and non-giant papers ($G = 0$) (Figure S3). We find that the $P(C)$ follows approximately a log-normal distribution for both giant and non-giant papers ($\mu \approx 34, \sigma \approx 1.1$ for giant papers and $\mu \approx 3.4, \sigma \approx 1.1$ for non-giant papers, respectively), consistent with the lognormal distribution of $P(C)$ reported in the literature (Radicchi, Fortunato, & Castellano, 2008). Yet, giant papers have significantly higher citations than non-giant papers, which is consistent with positive correlation between $G$ and $C$ (Figure 3a-c).

**Giant index distribution.** Does the giant index also follow a log-normal distribution? To answer this question, we calculate the distribution of giant index for papers published in different time periods. We find that $P(G)$ is overall quite stable across different time periods (Figure S4).

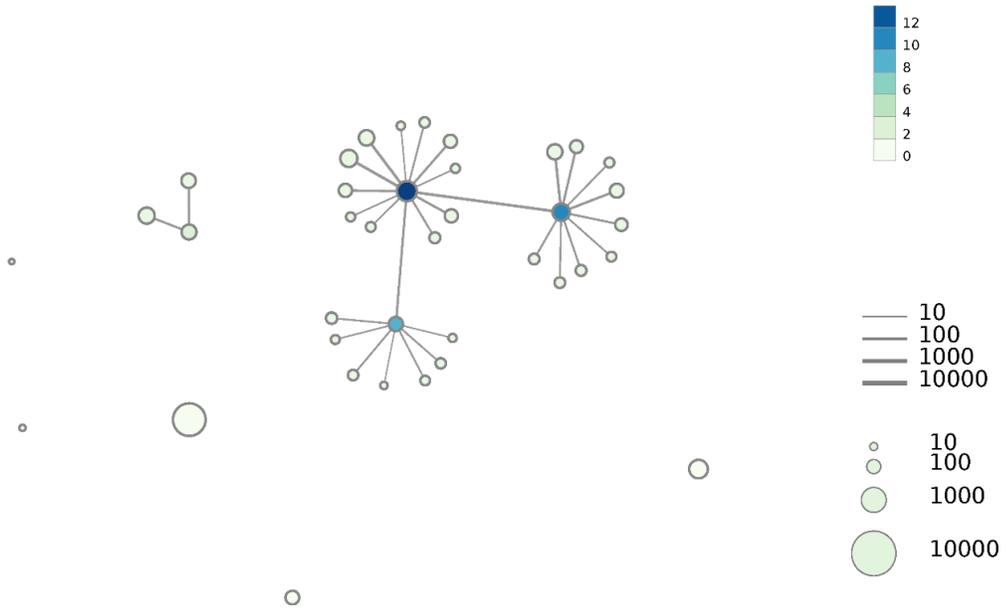

**Figure S1. The reference subnetwork for a research paper on CRISPR-Cas9[1]**. Nodes are references of the paper. Node color denotes the number of links, node size denotes the overall citation count, and link weight captures the co-citations between two nodes.

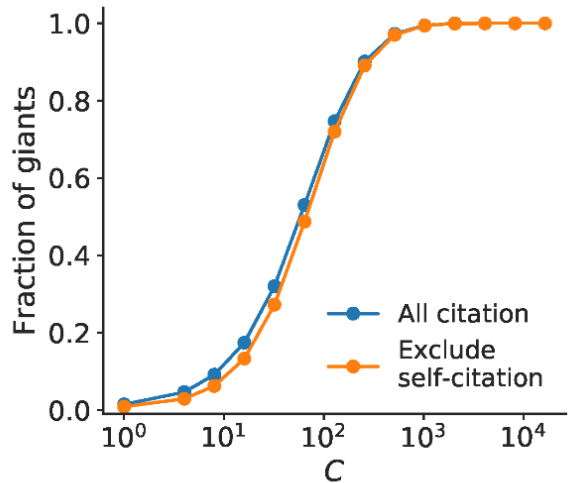

**Figure S2. The impact of self-citation.** The fraction of giants measured with and without self-citations for papers with different citation counts.

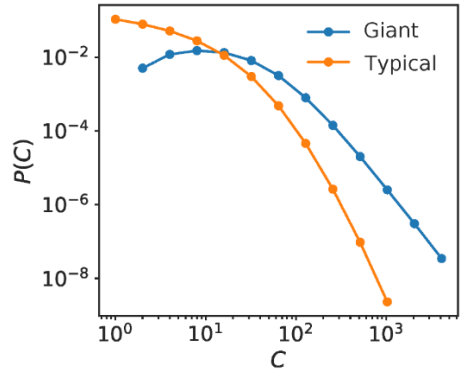

**Figure S3. The citation distribution $P(C)$ for giant papers ($G > 0$) and non-giant (typical) papers ($G = 0$)**

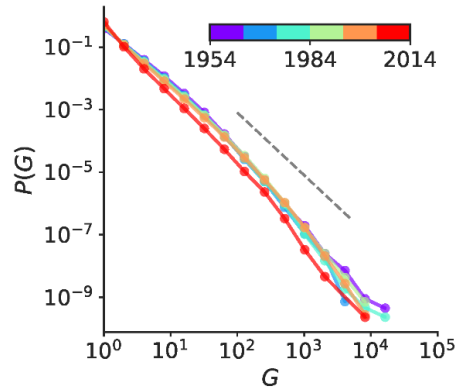

**Figure S4. The giant index distribution $P(G)$ for papers published each decade in the past 60 years.**